\documentclass[aps, prb, superscriptaddress, twocolumn, showpacs]{revtex4-1}

\usepackage[english]{babel}
\usepackage{hyperref}

\usepackage[pdftex]{graphicx}
\usepackage{color}

\usepackage{amsmath}
\usepackage{amssymb}
\usepackage{mathtools}
\usepackage{braket}

\usepackage[sort&compress]{natbib}

\newcommand{\un}[1]{\ensuremath{\,\mathrm{#1}}}

\newcommand{\fig}[1]{Figure~\ref{fig:#1}}
\renewcommand{\sec}[1]{Section~\ref{sec:#1}}
\newcommand{\tab}[1]{Table~\ref{sec:#1}}
\newcommand{\eq}[1]{Equation~(\ref{eq:#1})}
\newcommand{\lr}[1]{\ensuremath{\left( #1 \right)}}

\renewcommand{\Im}[1]{\ensuremath{\mathrm{Im} \left(#1\right)}}
\newcommand{\I}{\mathrm{i}}

\newcommand{\gm}{\gamma}

\newcommand{\Abs}[1]{\ensuremath{\left| #1 \right|}}
\newcommand{\Tr}[1]{\ensuremath{\mathrm{Tr}\left(#1\right)}}

\newcommand{\eps}{\epsilon}
\newcommand{\Sg}{\Sigma}
\newcommand{\sg}{\sigma}

\begin{document}

\title{Transport gap engineering by contact geometry in graphene nanoribbons:\\
  Experimental and theoretical studies on artificial materials}

\author{Thomas Stegmann}
\email{stegmann@icf.unam.mx}
\affiliation{Instituto de Ciencias F\'isicas, Universidad Nacional Aut\'onoma de M\'exico, 62210
  Cuernavaca, M\'exico}

\author{John A. Franco-Villafa\~ne}
\email{jofravil@ifisica.uaslp.mx}
\affiliation{CONACYT - Instituto de F\'isica, Universidad Aut\'onoma de San Luis Potos\'i, 78290 San
  Luis Potos\'i, M\'exico}
\affiliation{Instituto de Ciencias F\'isicas, Universidad Nacional Aut\'onoma de M\'exico, 62210
  Cuernavaca, M\'exico}

\author{Ulrich~Kuhl}
\affiliation{Universit\'{e} C\^{o}te d'Azur, CNRS, Institut de Physique de Nice, 06100 Nice, France}

\author{Fabrice Mortessagne}
\affiliation{Universit\'{e} C\^{o}te d'Azur, CNRS, Institut de Physique de Nice, 06100 Nice, France}

\author{Thomas H. Seligman}
\affiliation{Instituto de Ciencias F\'isicas, Universidad Nacional Aut\'onoma de M\'exico, 62210
  Cuernavaca, M\'exico}
\affiliation{Centro Internacional de Ciencias, 62210 Cuernavaca, M\'exico}

\begin{abstract}
  Electron transport in small graphene nanoribbons is studied by microwave emulation experiments and
  tight-binding calculations. In particular, it is investigated under which conditions a transport
  gap can be observed. Our experiments provide evidence that armchair ribbons of width $3m+2$ with
  integer $m$ are metallic and otherwise semiconducting, whereas zigzag ribbons are metallic
  independent of their width. The contact geometry, defining to which atoms at the ribbon edges the
  source and drain leads are attached, has strong effects on the transport. If leads are attached
  only to the inner atoms of zigzag edges, broad transport gaps can be observed in all armchair
  ribbons as well as in rhomboid-shaped zigzag ribbons. All experimental results agree qualitatively
  with tight-binding calculations using the nonequilibrium Green's function method.
\end{abstract}


\date{\today}

\maketitle

\section{Introduction \& Outline}
\label{sec:1}

\begin{figure}
  \centering
  (a) armchair ribbon\\
  \includegraphics[scale=0.39]{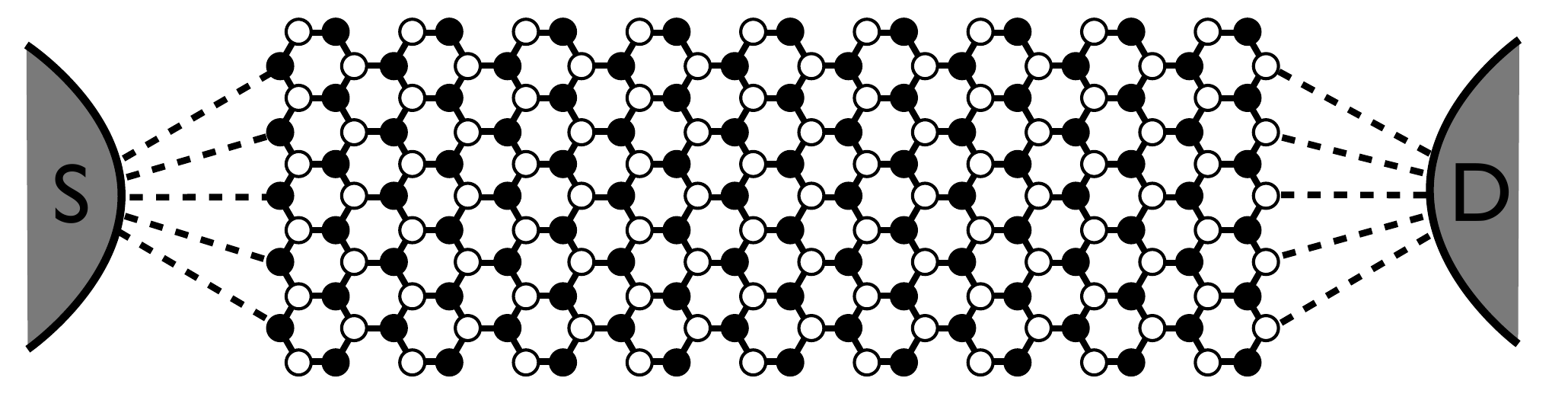}\\[1mm]
  \hspace{5mm}(b) zigzag ribbon \hspace{12mm} (c) zigzag-rhomboid ribbon\\
  \includegraphics[scale=0.38]{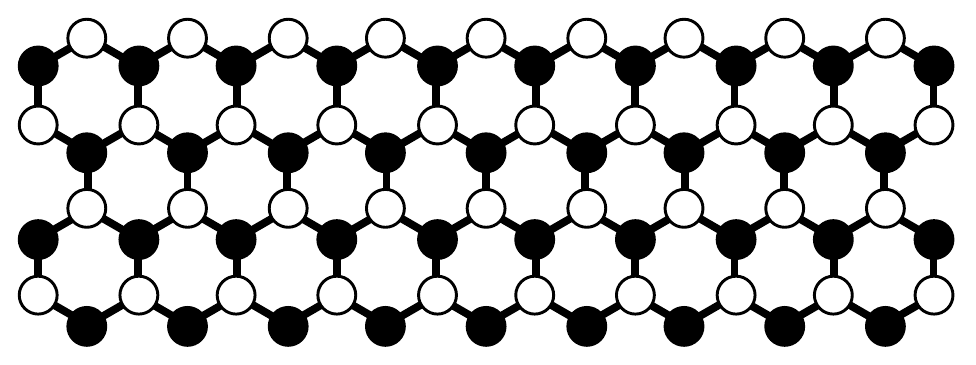} \hspace{3mm} \includegraphics[scale=0.38]{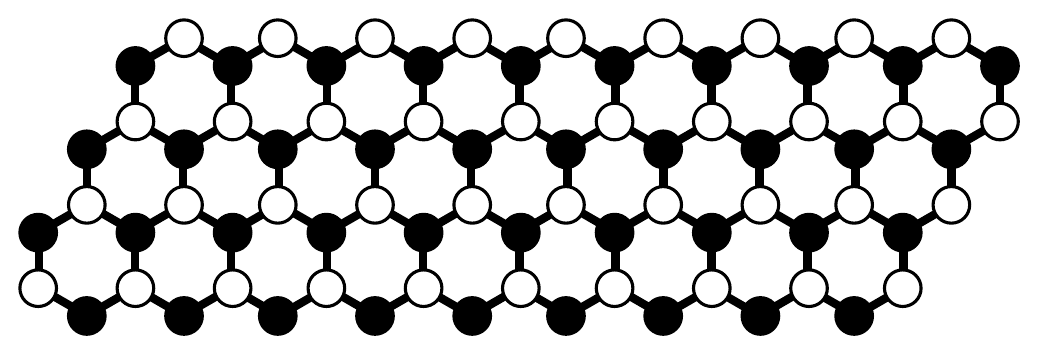}\\[1mm]
  (d) photo of the microwave experiment\\
  \includegraphics[scale=0.25]{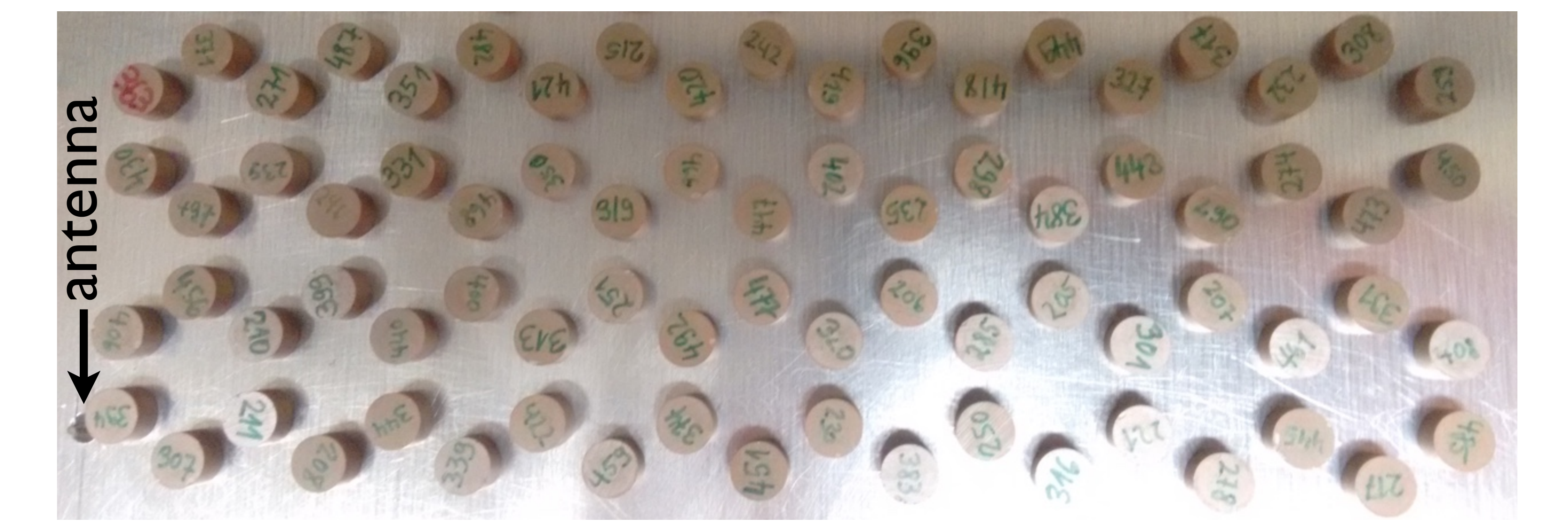}
  \caption{Electron transport through graphene nanoribbons from source (S) to drain (D) is studied
    by microwave emulation experiments and tight-binding calculations. Panel (a) shows an armchair
    ribbon of width 5, while zigzag and zigzag-rhomboid ribbons of width 3 are displayed in panel
    (b) and (c), respectively. The black and white color shading of the resonators indicate the two
    different sublattices to which the atoms belong. The contact geometry, i.e the way how source
    and drain are coupled to the ribbon (dashed lines), has important effects on the
    transport. Panel (d) shows a photo of the experimental setup for a zigzag ribbon of width 3. On
    the left hand side the antenna on the bottom plate is seen, whereas the antenna on the right
    hand side is mounted to the top plate (not shown).}
  \label{fig:1}
\end{figure}

Nowadays, graphene is one of the most studied materials in condensed matter physics because of its
various exceptional properties and their technical applications, see Refs.~\onlinecite{Geim2009,
  CastroNeto2009, Avouris2010, Novoselov2012, Katsnelson2012, Ferrari2015, Aoki2014,Torres2014} for
an overview. One of the most remarkable features is that graphene has a linear dispersion relation
at the Dirac points, which lets the electrons behave as relativistic, massless, charged
fermions. The high mobility of the charge carriers, coming from the special dispersion relation at
the Fermi energy, makes graphene very promising for new electronic devices. However, the absence of
a band-gap in graphene inhibits to substitute nowadays silicon-based semiconductor technology by
graphene.\cite{Schwierz2010} One approach to open a band-gap in graphene is to use nanoribbons,
i.e. small stripes of graphene, see \fig{1}. On one hand, this approach has the advantage that the
rather small size of graphene nanoribbons may lead to a high miniaturization and integration of
these devices. On the other hand, it has the disadvantage that it is still challenging to produce
nanoribbons of well controlled size and geometry, although there is promising progress, see for
example Refs.~\onlinecite{Cai2010, Koch2012, Ruffieux2012, Chen2013b, Cai2014, Chen2015,
  Kimouche2015, Ruffieux2016}. Moreover, connecting nanoribbons to leads, where electrons are
injected and extracted, is experimentally demanding. Note also that graphene nanoribbons are
predicted to operate as valley filters, see for example Ref.~\onlinecite{Rycerz2007,
  Nakabayashi2009, Gunlycke2011}. The so-called valleytronics, where the pseudospin of the charge
carriers is used, may lead to new electronic devices, which do not have an analog in silicon-based
electronics.

Recently, it has been shown that a tight-binding model of graphene and polyacetylene can be emulated
by microwave experiments.\cite{Bellec2013, Bellec2013b, Barkhofen2013, Stegmann2017} Such
experiments are well controlled and easy to perform (in comparison to experiments with real graphene
or polyacetylene) and hence, offer a versatile tool to investigate in detail the properties of these
systems. In this article, we study the ballistic single-electron transport through small graphene
nanoribbons by microwave transmission experiments. Our measurements are supported by tight-binding
calculations using the nonequilibrium Green's function (NEGF) method.\cite{Datta1997, Datta2005,
  Lewenkopf2013}

Graphene nanoribbons have two elementary edge structures, the zigzag and the armchair shape, see for
example the horizontal edges in \fig{1} (a) and (b), respectively. Edge deformations are not
considered here.\cite{Hawkins2012} Studies of graphene nanoribbons\cite{Fujita1996, Nakada1996,
  Brey2006, Wakabayashi2009, Wakabayashi2010} predict that armchair ribbons of width $3m+2$ with
integer $m$ are metallic and otherwise semiconducting, i.e. they show a broad band-gap at the Dirac
point.\footnote{Studies using density function theory \cite{Son2006, Motta2012} indicate that also
  in the metallic armchair ribbons a narrow band-gap can be observed. However, these correlation
  effect go beyond the present study.} Zigzag ribbons are predicted to be metallic for all ribbon
widths. First samples of small graphene nanoribbons have been synthesized recently\cite{Cai2010,
  Koch2012, Ruffieux2012, Chen2013b, Cai2014, Chen2015, Kimouche2015, Ruffieux2016} indicating the
predicted behavior. Here, we present emulation experiments of the electronic transport through small
nanoribbons with specific edges and atomically precise connections to source and drain leads. We do
not only provide further evidence to the predicted behavior but study also the effect of the contact
geometry, which determines to which sites at the edges of the ribbons leads are attached. We show
that by tuning the contact geometries, broad transport gaps can be induced in graphene nanoribbons
with armchair and zigzag edges independent from their actual width. Contact effects on the transport
in graphene nanoribbons have been addressed only rarely. Square lattices have been attached to the
honeycomb lattice of graphene ribbons,\cite{Blanter2007, Schomerus2007, Zhang2010, Zhang2011b,
  Mochizuki2009, Mochizuki2010, Pieper2013} which for example induces in zigzag ribbons of even
width a transport gap, while ribbons of odd width remain metallic. Similar even-odd parity effects
can be observed also with respect to the total length of the ribbon.\cite{Li2008} The case where
leads are attached to small \cite{Maiti2009, Li2009} and larger \cite{Konopka2015} nanoribbons has
also been studied.

The paper is organized as follows. In \sec{2}, we describe the studied systems and explain briefly
the used experimental and theoretical methods. Our results are presented and discussed in
\sec{3}. Conclusions and an outlook can be found in \sec{4}.

\section{System \& Methods}
\label{sec:2}

We consider graphene nanoribbons of the types shown in \fig{1}, namely an armchair ribbon (a), a
zigzag ribbon (b) and a zigzag-rhomboid ribbon (c). The naming of the ribbons is due to the shape of
their horizontal edges and, in the case of \fig{1} (c), due to the overall shape of the
ribbon. Because of experimental limitations, the length of the ribbons is kept as shown in \fig{1},
while their width is varied. However, calculations have been performed also for larger systems,
giving qualitatively the same results discussed below. Leads through which electron (or microwaves)
are injected and extracted, are attached to the atoms at the left and right edges, see the dashed
lines in \fig{1} (a).

\subsection{Microwave experiment}
\label{sec:2-1}

Applying the techniques, developed to investigate the band structure of graphene\cite{Bellec2013,
  Bellec2013b} and to emulate relativistic systems\cite{Franco2013, Sadurni2013} as well as
molecular systems\cite{Stegmann2017}, we have performed analogous experiments to study the coherent
transport in graphene nanoribbons.

A set of identical dielectric cylindrical resonators ($5\un{mm}$ height, $4\un{mm}$ radius,
refractive index $n\approx 6$) is placed between two metallic plates. A photo of the experimental
setup without the top plate is shown in \fig{1} (d). The individual resonators have an isolated
resonance at $\nu_0'=6.655 \pm 0.005 \un{GHz}$, corresponding to the lowest transverse electric (TE)
mode. We restrict our investigation to frequencies around $\nu_0'$, where each resonator contributes
only one resonance. The nearest neighbor distance between the center of the resonators is
$d_1=12.0 \un{mm}$. The dielectric resonators play the role of the carbon atoms in the ribbons,
while the electromagnetic waves corresponds to the wave function of the electron. A detailed
description of the experimental setup can be found in Ref.~\onlinecite{Bellec2013b}. As the antennas
are positioned always close to or above a single resonator they couple to the closest resonator
only, thus measuring the transmission from an individual resonator on the left hand side to another
individual resonator at the right hand side. By changing the antenna positions all combinations of
transmission between edge resonators are measured. Due to the weak coupling of the antennas the
total transmission is then given by summing up all contributions, see \sec{2-2}. As the system is
time-reversal invariant the transmission is reciprocal, i.e. the transport in both directions is the
same.

\subsection{Tight-binding transport calculations}
\label{sec:2-2}

Theoretically, the graphene nanoribbons are described by the tight-binding Hamiltonian
\begin{equation}
  \label{eq:1}
  H= \sum_{\Abs{i-j} \leq 3\text{nn}} t_{\Abs{i-j}} \ket{i}\bra{j}.
\end{equation}
The coupling parameters $t_{\Abs{i-j}}$, which have been obtained by fitting our calculations to the
experimental data, are given in \tab{1}. In the sum in \eq{1}, interactions up to third nearest
neighbors (3nn) are taken into account. Interactions to higher nearest neighbors can be safely
ignored due to the large distance of these sites and the additional screening by the closer sites.

\begin{table}[h]
  \centering
  \begin{tabular}{c|c|c|c}
    & $t_1$ & $t_2$ & $t_3$\\
    \hline
    \begin{minipage}{13mm}
      \centering
      \includegraphics[scale=0.26]{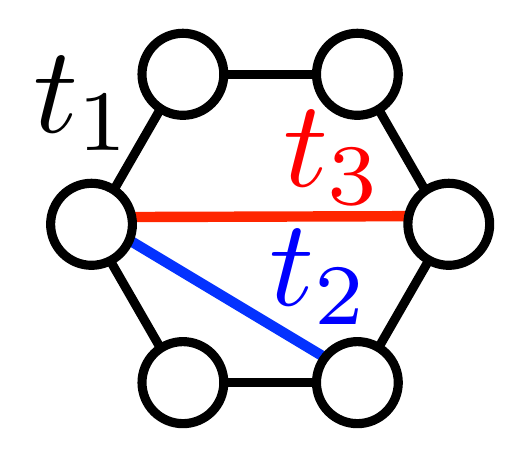}
    \end{minipage} & 39.5 & \,5.0\, & \,3.0\,
  \end{tabular}
   \caption{Coupling parameters $t_i$ (in MHz) up to third nearest neighbors. The three parameters
     have been obtained by fitting our calculations to the experimental data.}
  \label{tab:1}
\end{table}

From now on we will use the normalized frequency $\nu= \nu' -\nu_0' - 3t_2$, where the Dirac point
is located theoretically\cite{Munarriz2014} in the center of the transmission band (i.e. at
$\nu=0$).

The electron transport through the nanoribbons is calculated by means of the nonequilibrium Green's
function (NEGF) method. In the following, we briefly summarize the essential equations. Details can
be found in Refs.~\onlinecite{Datta1997, Datta2005, Lewenkopf2013}. The Green's function of the
chain is defined as
\begin{equation}
  \label{eq:2}
  G(\nu)= \Big[\nu -H -\Sg_S -\Sg_D -\Sg_{\text{abs}} -\Sg_{\text{dis}} \Big]^{-1},
\end{equation}
where $\nu$ is the electron energy corresponding to the microwave frequency in the experiment.

The effect of the source and drain leads by which electrons are injected and extracted, is described
by the self-energies
\begin{equation}
  \label{eq:3}
  \Sg_{S}= -\I \eta \sum_{i \in S}\ket{i}\bra{i}, \quad
  \Sg_{D}= -\I \eta \sum_{i \in D}\ket{i}\bra{i}.
\end{equation}
The coupling strength $\eta= 0.6 \un{MHz}$ is adjusted to the experiment. The sums are over the
sites where leads are attached, see for example in \fig{1} (a) the sites which are connected by
dashed lines to the source and drain, respectively. The sites connected to leads are also indicated
in the insets of figures \ref{fig:2}, \ref{fig:4} and \ref{fig:5} by the black sites. The
self-energies show no coherences, i.e. no off-diagonal elements, which is justified because in the
experiment the antennas are coupled only weakly to the resonators.

Absorption, which is present in the experiment, is modeled by the imaginary potential (or
self-energy)
\begin{equation}
  \label{eq:4}
  \Sg_{\text{abs}}= -\I \gm(W) \sum_{i=1}^N \ket{i} \bra{i},
\end{equation}
where $N$ is the total number of sites of the ribbon. We obtain best agreement between the
experimental data and the calculations using for the armchair ribbons a linear decay of the
absorption $ \frac{\gm(W)}{\text{MHz}}= 0.99 -0.06W $, where $W$ is the width of the ribbon measured
in multiples of the hexagonal cell size. For the zigzag and zigzag-rhomboid ribbons, we use a
constant absorption $ \gm(W)= 0.99 \un {MHz} $.

In the experiment, some degree of disorder cannot be avoided completely due to the uncertainty of
the resonance frequency of the resonators and the uncertainty of their positions. In the
calculations, disorder is taken into account by a random potential (or self-energy)
\begin{equation}
  \label{eq:6}
  \Sg_{\text{dis}}= \sum_{i=1}^N \eps_i \ket{i} \bra{i},
\end{equation}
where the $\eps_i$ are chosen from a Gaussian distribution which is cut at its full width half
maximum, which corresponds approximately to the experimentally observed distribution and the used
selection rule. We consider the standard deviation $\sg=10 \un{MHz}$ and an ensemble of $10^2$
realizations.

The transmission between source and drain is then given by
\begin{equation}
  \label{eq:7}
  \begin{aligned}
    T(\nu)&= 4\Tr{\Im{\Sg_S}G\,\Im{\Sg_D}G^\dagger}\\
    &= {\textstyle \sum_{i \in S, j \in D}} \, T_{ij}.
  \end{aligned}
\end{equation}
where $T_{ij}(\nu)=4 \eta^2 \Abs{G_{ij}}^2$ is the transmission between an individual site $i$ at
the left end of the ribbon to another site $j$ at the right end. As discussed in \sec{2-1}, these
functions $T_{ij}$ are measured in the microwave experiment. Note that in the last step in \eq{7},
we have used the fact that the self-energies in \eq{3} describing the effect of source and drain are
sparse matrices with only some non-vanishing entries on their diagonals.

In order to understand the transport properties, we will also calculate the local density of states
(LDOS)
\begin{equation}
  \label{eq:8}
  D(\nu)= \text{Diag}\lr{G\left[\Sg_S+\Sg_D\right]G^\dagger}.
\end{equation}
Due to the weak coupling of the leads to the nanoribbon, see \eq{3}, the LDOS is very similar to the
eigenstates of the closed Hamiltonian in \eq{1}, which are near to the considered frequency. Contact
induced states\cite{Golizadeh-Mojarad2009} have only minor effects here.

We would like to emphasize that the parameters adjusted to match the experiment and the numerics are
a minimal and well defined set. Each parameter impose specific and distinct features on the measured
spectra. The resonance frequency $\nu'_0$ and the disk couplings ($t_1$, $t_2$, $t_3$) define band
center, band width, and the asymmetry of the two bands\cite{Bellec2013b}. The antenna coupling
$\eta$ determines mainly the resonance depth, whereas the absorption $\gamma(W)$ is mainly related
to the smoothing. The disorder strength $\eps$ takes into account fluctuation in a statistical
sense, therefore a perfect agreement between experiment and numerics cannot be expected.

\section{Results \& Discussion}
\label{sec:3}

\subsection{Armchair ribbons}
\label{sec:3-1}

\begin{figure*}
  \centering
  \includegraphics[scale=0.415]{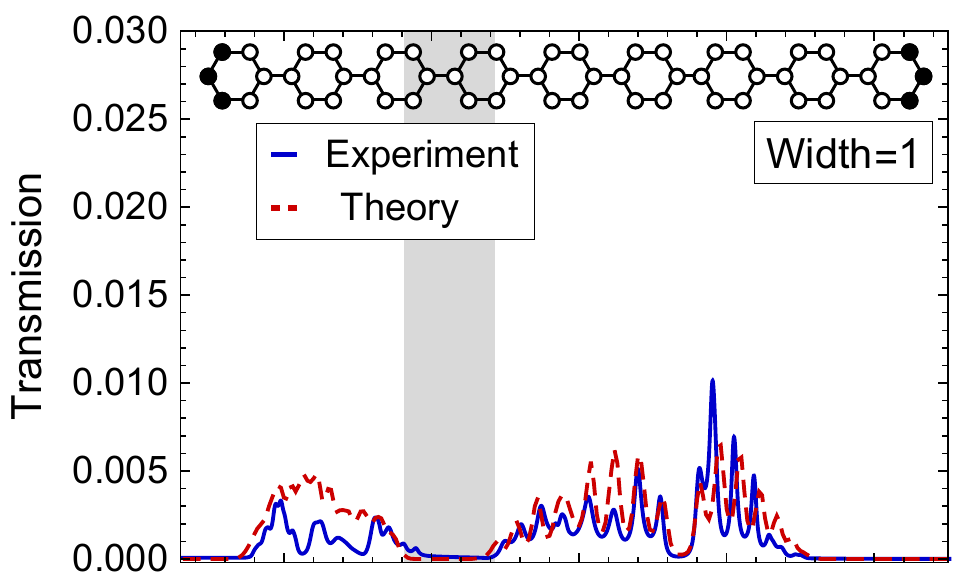}
  \includegraphics[scale=0.415]{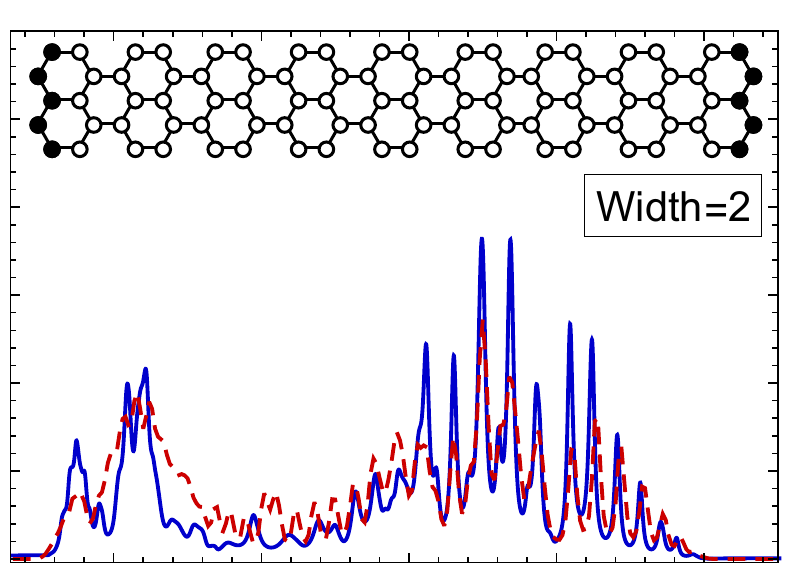}
  \includegraphics[scale=0.415]{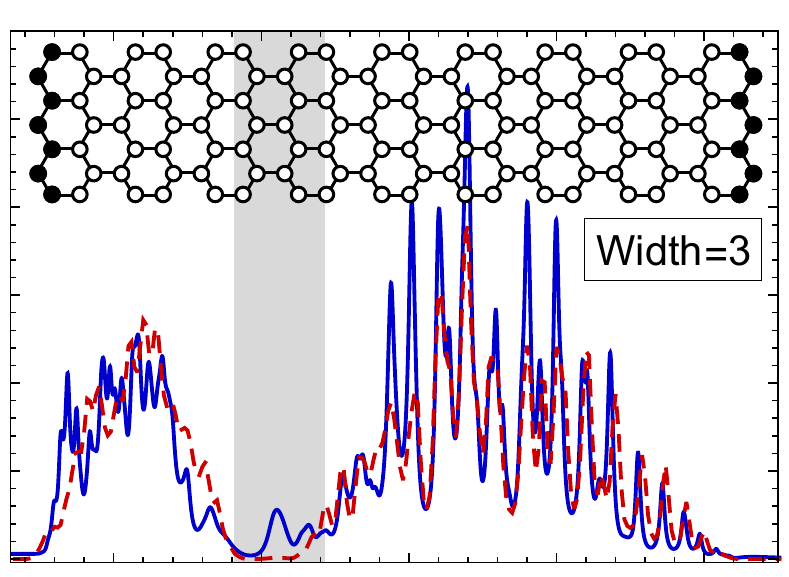}
  \includegraphics[scale=0.415]{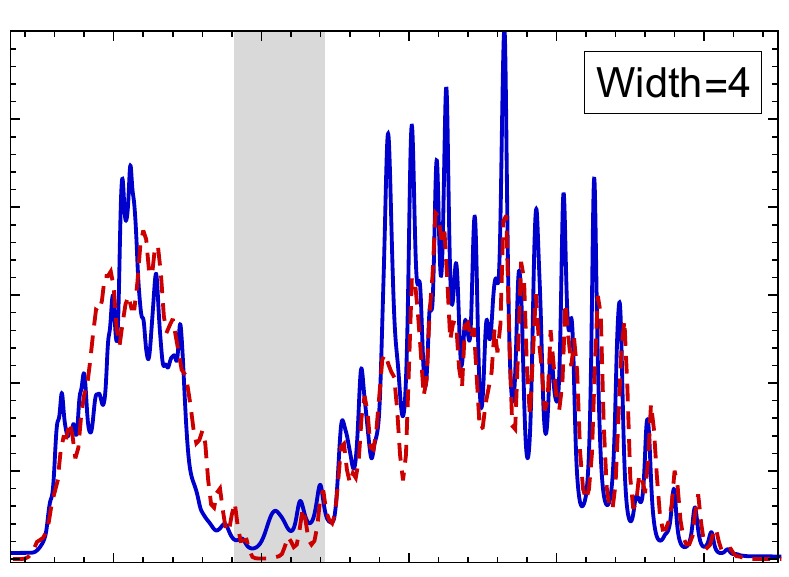}
  \includegraphics[scale=0.415]{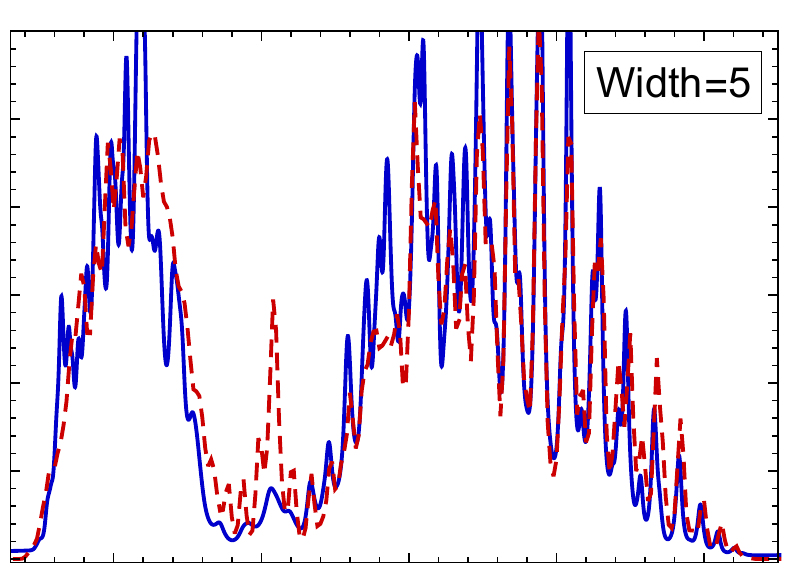}\\[2mm]
  \includegraphics[scale=0.415]{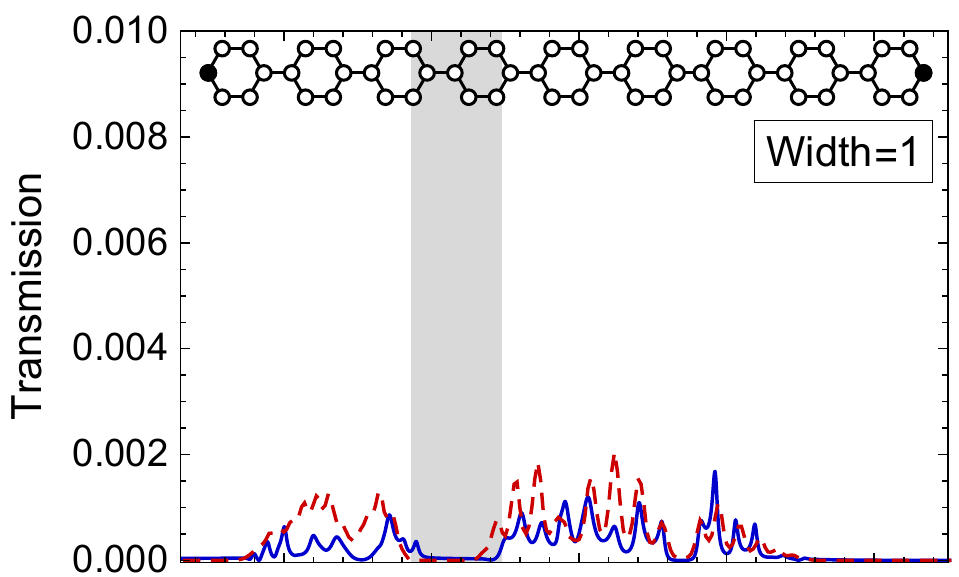}
  \includegraphics[scale=0.415]{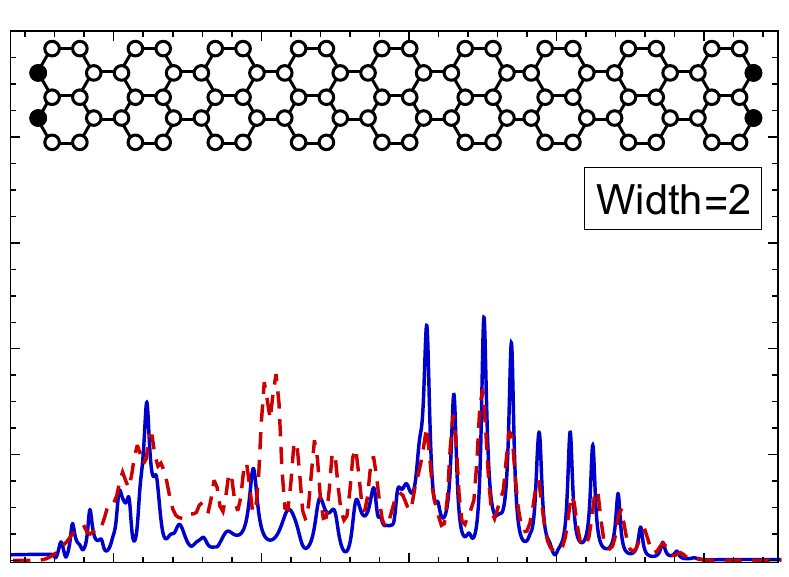}
  \includegraphics[scale=0.415]{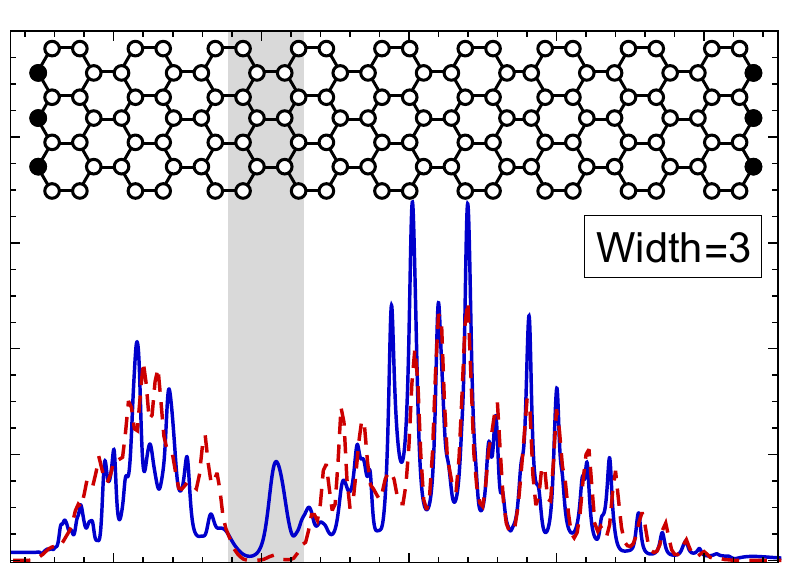}
  \includegraphics[scale=0.415]{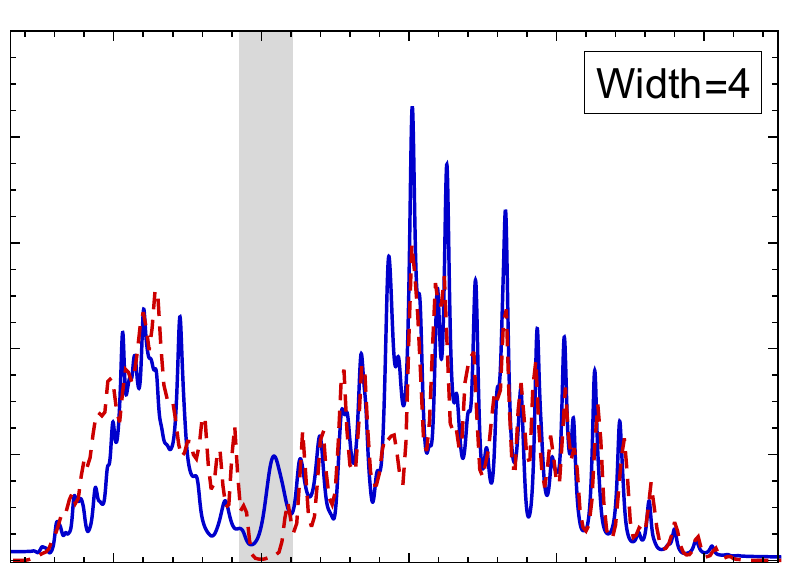}
  \includegraphics[scale=0.415]{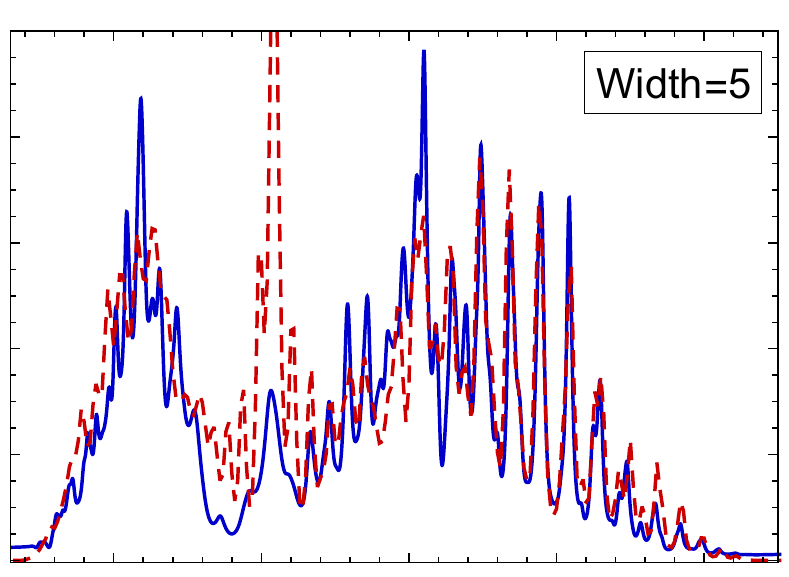}\\[2mm]
  \includegraphics[scale=0.415]{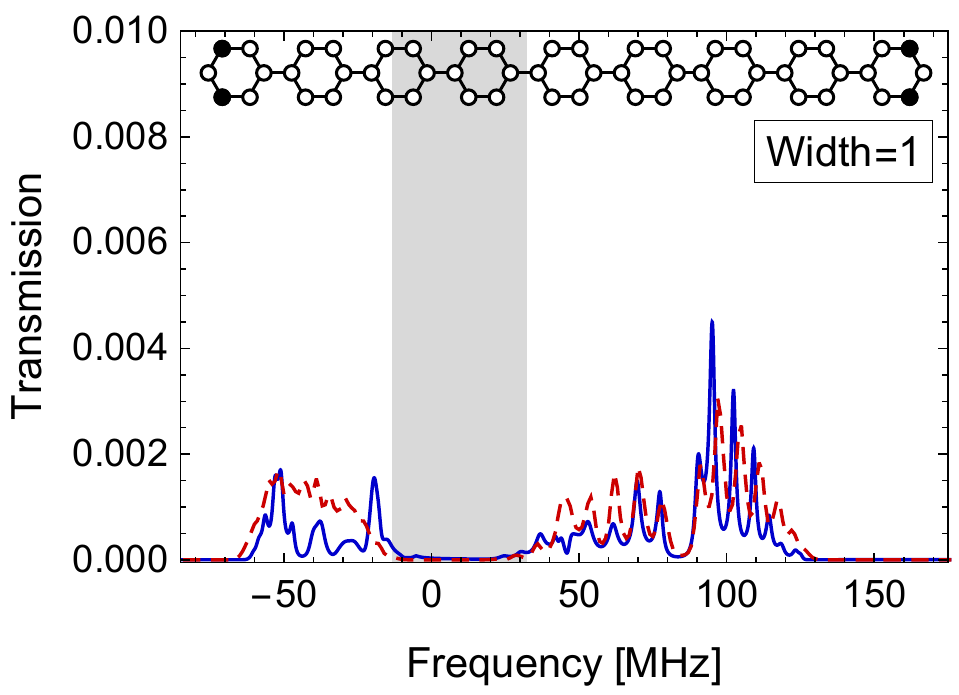}
  \includegraphics[scale=0.415]{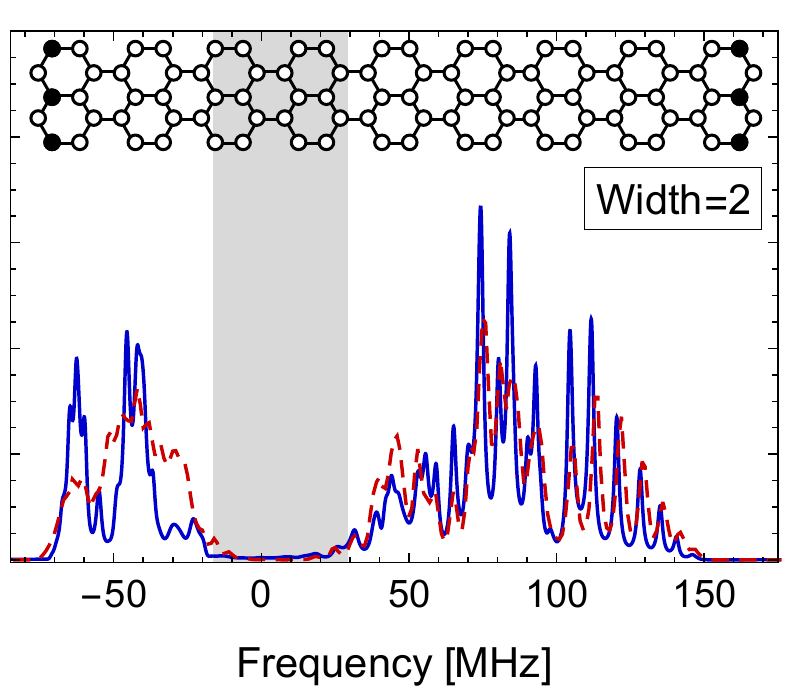}
  \includegraphics[scale=0.415]{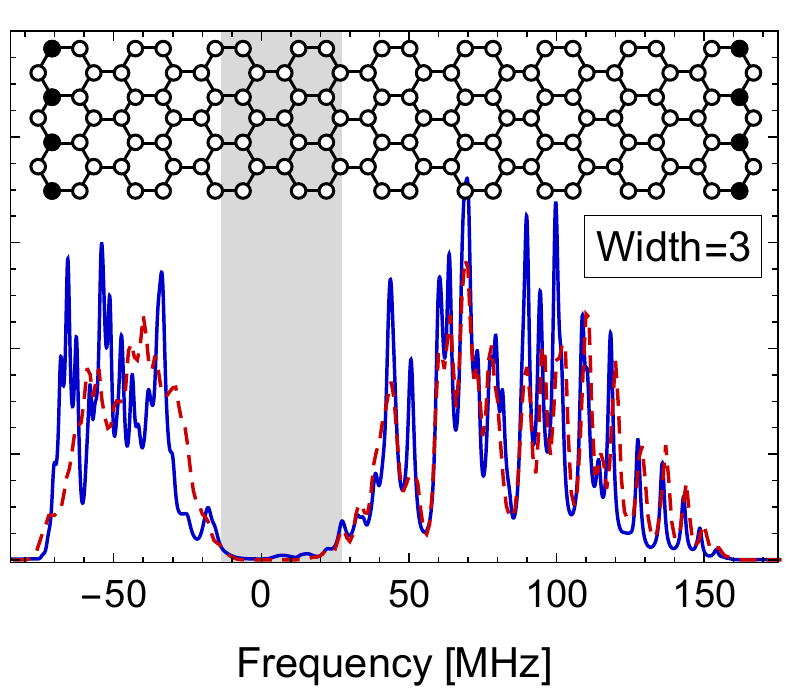}
  \includegraphics[scale=0.415]{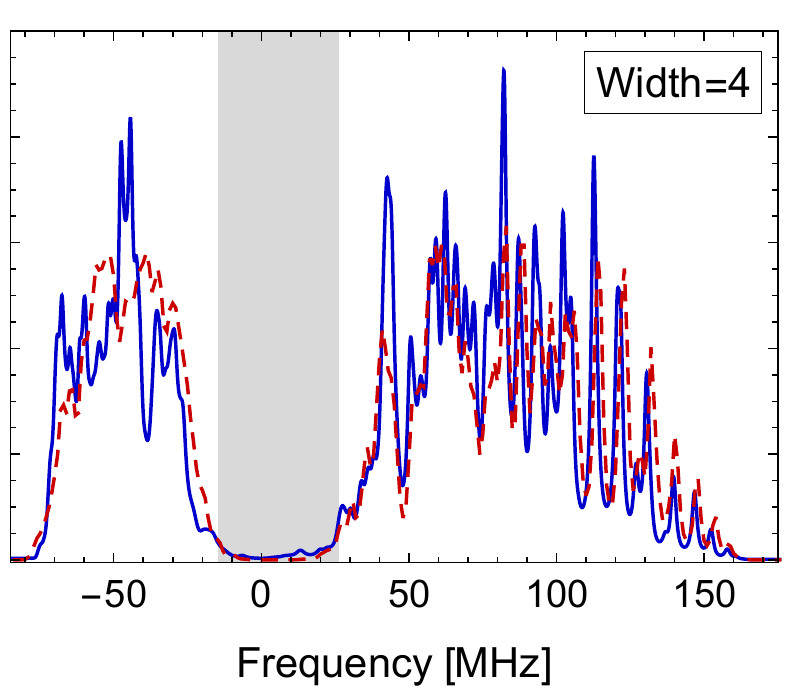}
  \includegraphics[scale=0.415]{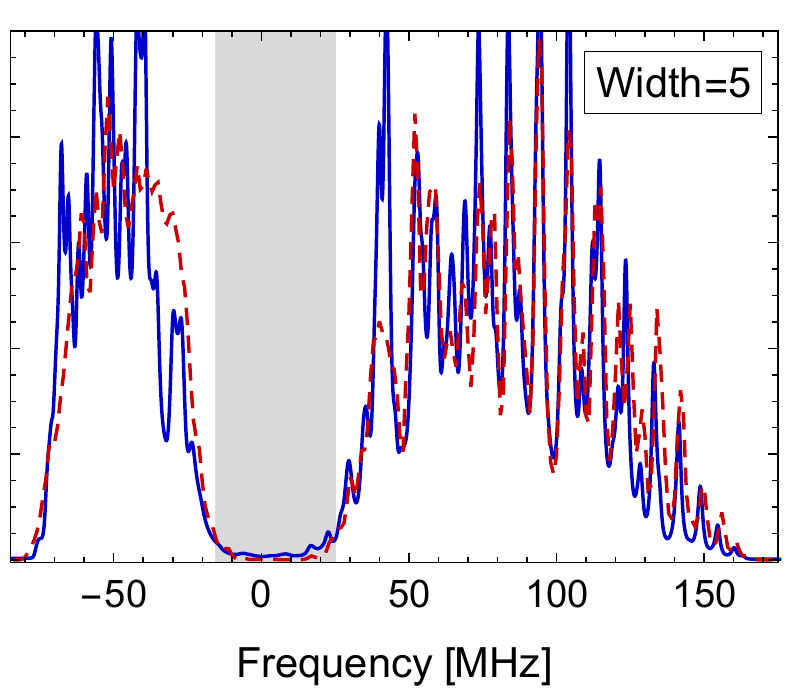}
  \caption{Transmission through armchair graphene nanoribbons of length $9$ and for increasing
    widths (from left to right). The ribbons up to a width of $3$ are sketched in the insets. The
    leads (or antenna) through which electrons (or microwaves) are injected and extracted are
    connected to the black sites at the edges of the ribbons. Experimental data is shown by
    blue-solid lines, while our tight-binding Green's function calculations are indicated by
    red-dashed lines. First row: Connecting leads to all atoms on the zigzag edges, we confirm that
    the ribbons of width $2$ and $5$ are metallic, while otherwise the ribbons show a transport gap
    (gray shaded regions) around $\nu=0$. Second row: Connecting only the outer atoms on the zigzag
    edges, see the black marked atoms in the insets, the differences between the metallic (width $2$
    and $5$) and semiconducting (width $1$, $3$ and $4$) ribbons become more pronounced. Third row:
    Connecting only the inner atoms of the zigzag edges to leads, we find a broad transport gaps for
    all ribbon widths.}
  \label{fig:2}
\end{figure*}

The transmission through graphene armchair ribbons for various widths and contact geometries is
shown in \fig{2}. Experimental data are indicated by the blue-solid curves while our tight-binding
calculations are highlighted by the red-dashed curves. The narrow ribbons are sketched in the
insets, where the black shaded sites are connected to leads.

We observe that in all ribbons the conductivity decreases when the Dirac point around $\nu=0$ is
approached. In the first row the leads are attached to all atoms at the confining zigzag edges to
the left and right. For ribbons of width $3m+2$ with integer $m$ (i.e. widths 2 and 5 in \fig{2}),
the conductivity around the Dirac point is low but finite. These ribbons are metallic in the whole
transmission band ranging approximately form $-80$ to $160 \un{MHz}$. For ribbons of other widths
(i.e. widths 1, 3 and 4 in \fig{2}) the conductivity around the band-center approaches zero. A
transport gap (gray shaded regions) opens in these ribbons, which hence are semiconducting. We
therefore provide clear experimental evidence of this theoretically predicted \cite{Fujita1996,
  Nakada1996, Brey2006, Wakabayashi2009, Wakabayashi2010, CastroNeto2009, Heikkilae2013} behavior of
graphene armchair ribbons. In general, the tight-binding calculations agree well with the
experimental data. In the upper part of the conduction band (from $0$ to $160 \un{MHz}$) almost all
resonance peaks coincide. In the lower part (from $-80$ to $0 \un{MHz}$) the agreement is not that
perfect but the general trend of the experimental data is reproduced also there. Differences between
experiment and theory are more notorious in the gap than outside. In particular for certain ribbon
width the experiment shows peaks which are not present in our calculations and which we attribute to
localized states enhanced by irregularities of the experimental setup.

The differences between the metallic and semiconducting armchair ribbons become more pronounced if
the leads are attached only to the outer atoms of the confining zigzag edges, see the second row in
\fig{2}. The transport through the ribbons changes drastically, if the leads are attached only to
the inner atoms, as shown in the third row of \fig{2}. In this case a clearly pronounced transport
gap is observed for all armchair ribbons, independent from their actual width.

\begin{figure}
  \centering
  $\nu=10 \un{MHz}$\\
  \includegraphics[scale=0.65]{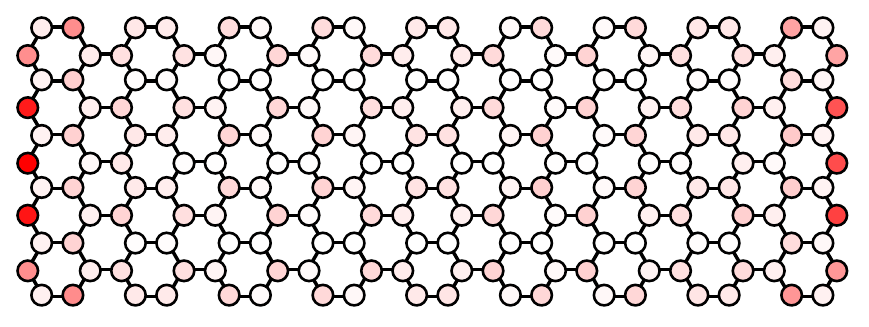}\\
  \includegraphics[scale=0.34]{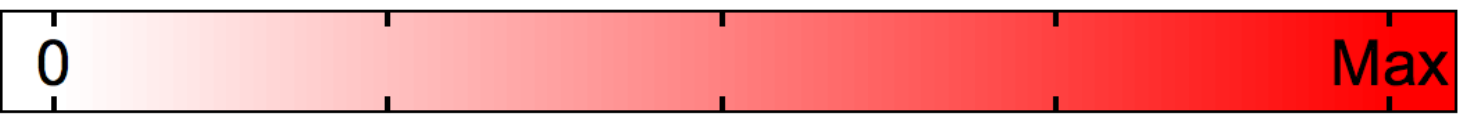}
  \caption{Calculated LDOS $D(\nu)$ in the armchair ribbon indicated by the color-shading of the
    resonators. Close to the Dirac point, the LDOS vanishes on the inner atoms of the confining
    zigzag edges to the left and right, while it is non-zero on the outer atoms. Note that leads are
    attached to all sites at the confining zigzag edges to the left and right.}
  \label{fig:3}
\end{figure}

\begin{figure*}
  \centering
  \vspace*{-3mm}
  \includegraphics[scale=0.415]{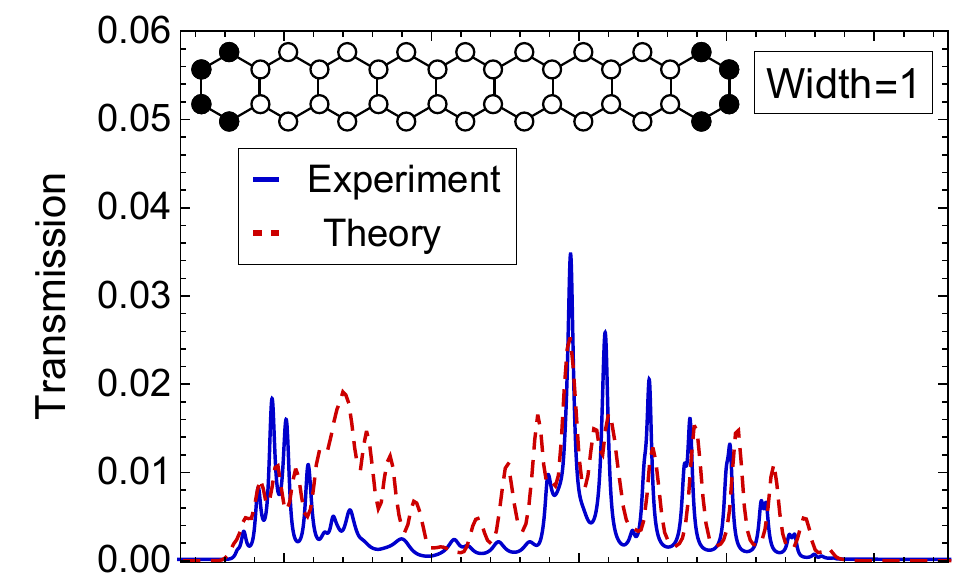}
  \includegraphics[scale=0.415]{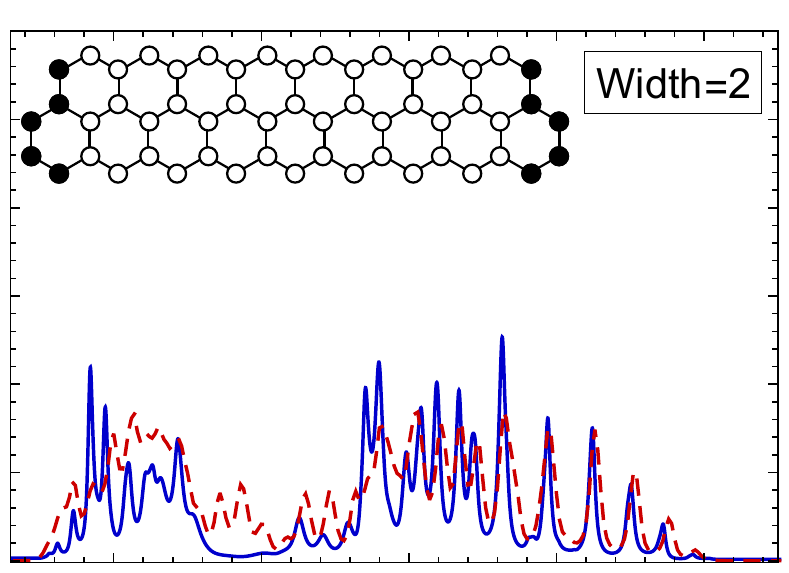}
  \includegraphics[scale=0.415]{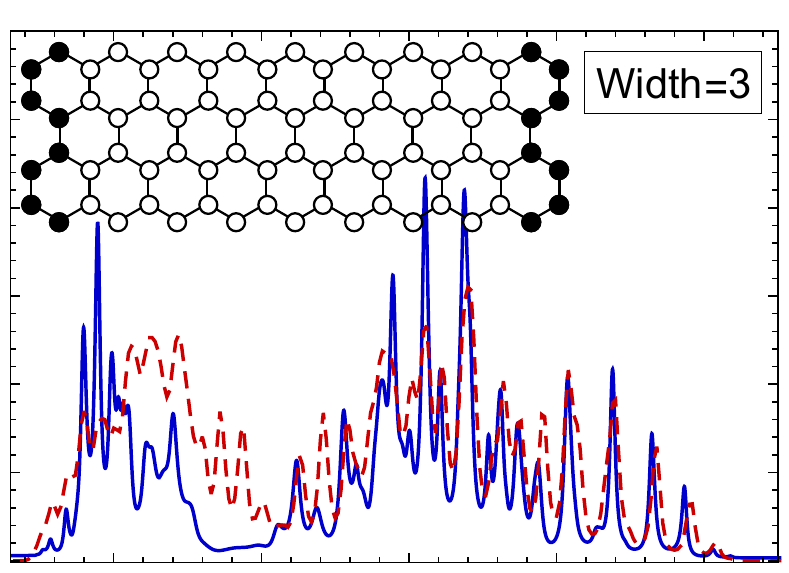}
  \includegraphics[scale=0.415]{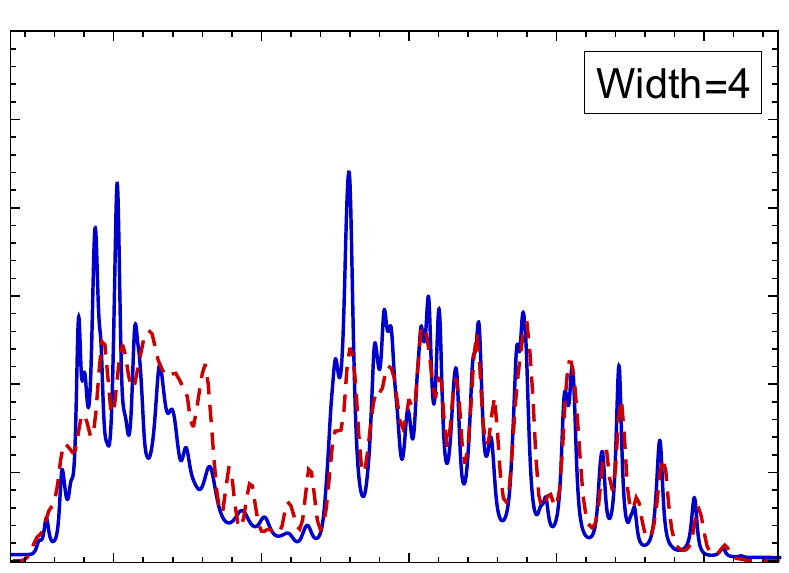}
  \includegraphics[scale=0.415]{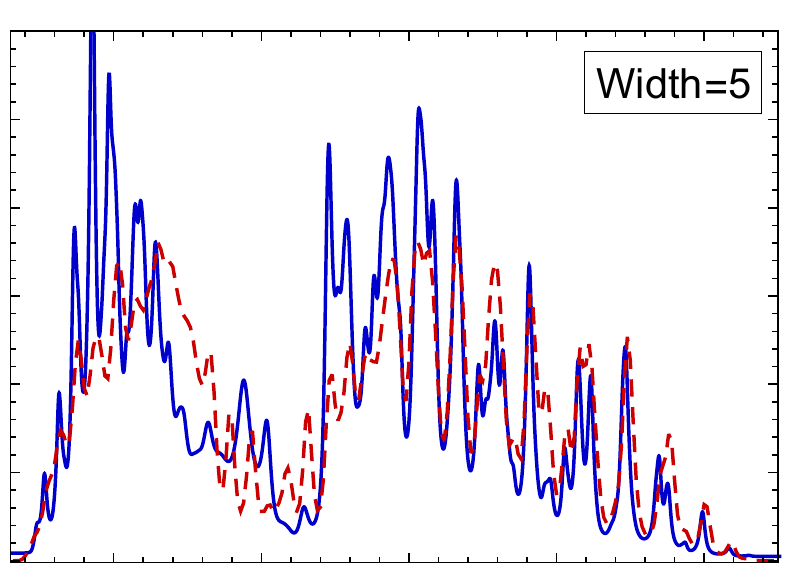}\\[2mm]
  \includegraphics[scale=0.415]{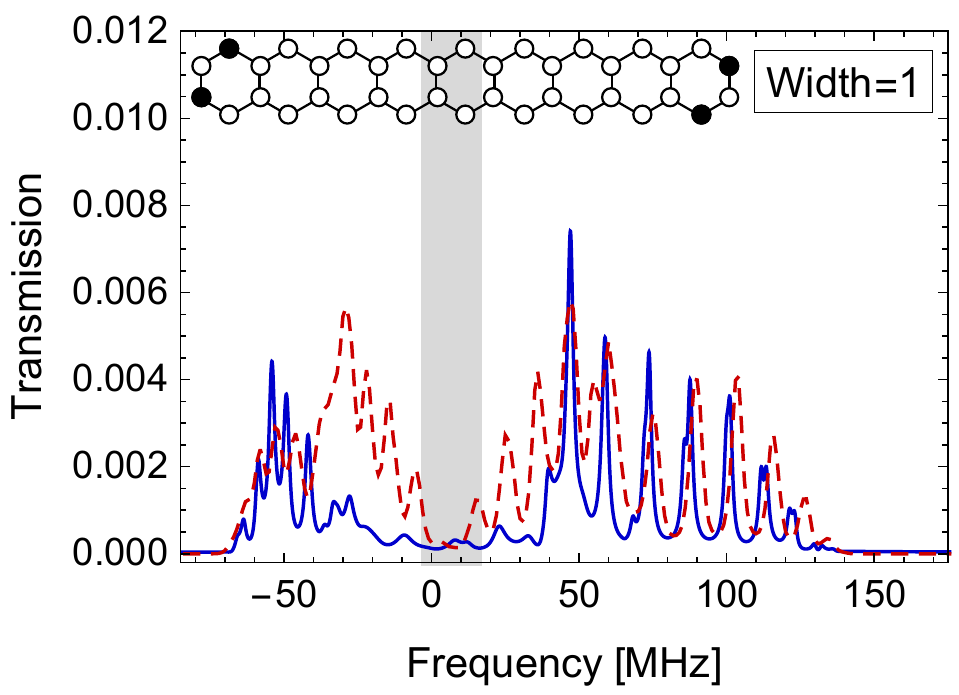}
  \includegraphics[scale=0.415]{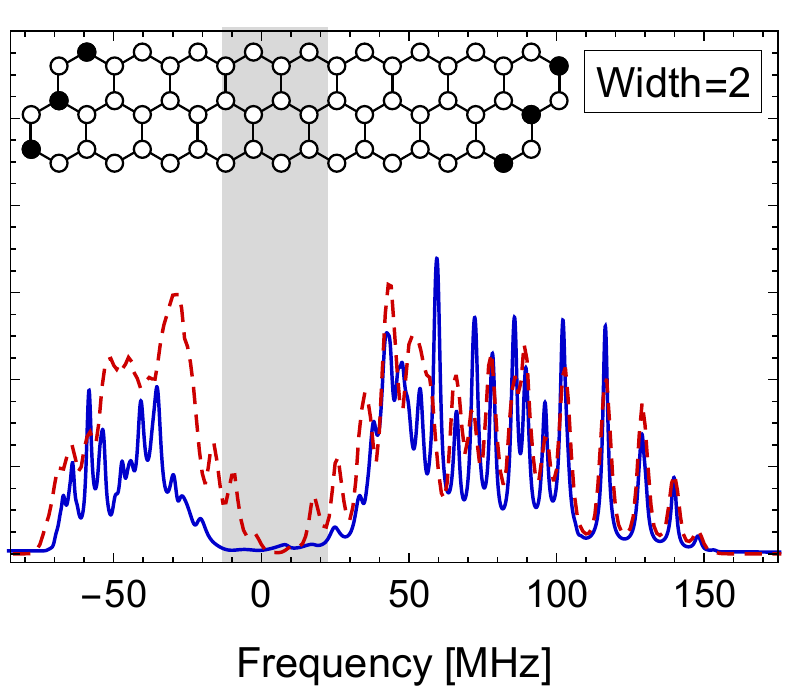}
  \includegraphics[scale=0.415]{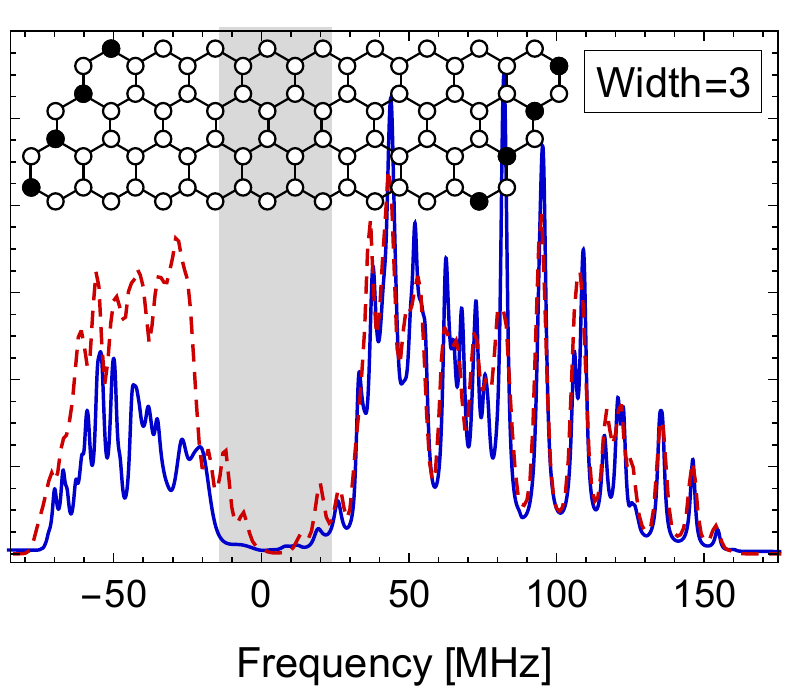}
  \includegraphics[scale=0.415]{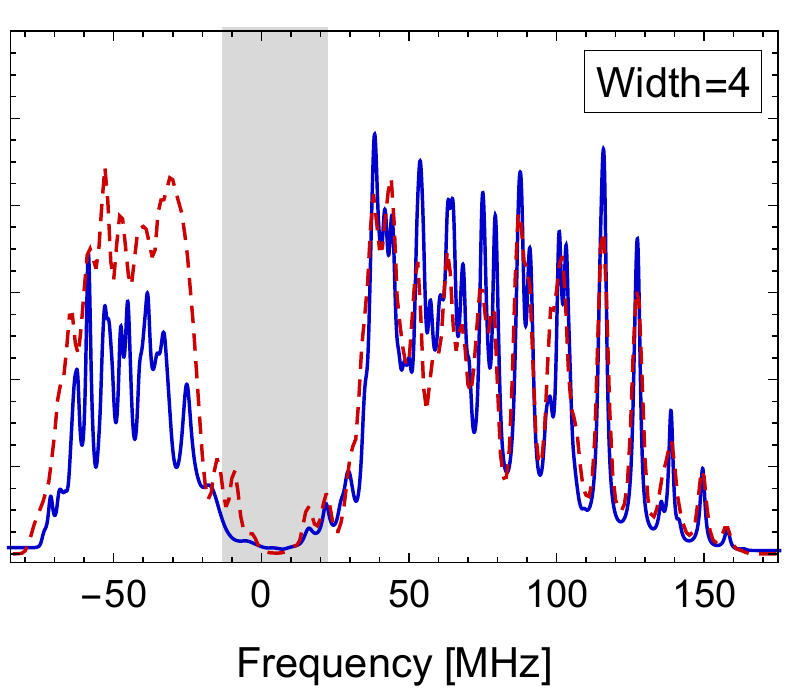}
  \includegraphics[scale=0.415]{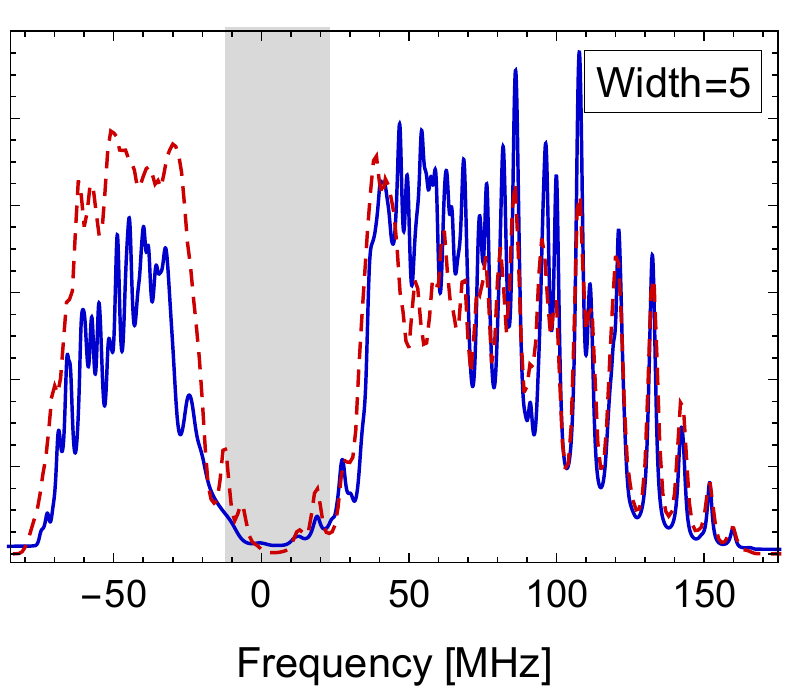}
  \caption{Transmission through zigzag and zigzag-rhomboid ribbons of length $9$ and increasing
    width (from left to right). The more narrow ribbons are sketched in the insets. The leads are
    attached to the black sites. Top row: Connecting leads to all atoms at the confining armchair
    edges to the left an right, we observe that all zigzag ribbons are metallic. Bottom row: If
    leads are connected to the inner atoms of a zigzag-rhomboid ribbon, broad transmission gaps
    (gray shaded regions) can be observed for all ribbons.}
\label{fig:4}
\end{figure*}

In order to understand the effects of the contact geometry on the transport, we show in \fig{3} the
local density of states (LDOS), which has been calculated by means of \eq{8}, near to $\nu=0$ where
the transport gap appears. At the zigzag edges to the left and right, where the leads are attached,
the LDOS is localized on the outer atoms, whereas it vanishes on the inner atoms. This property can
be observed for all frequencies within the gap. Thus, the inner atoms on zigzag edges are
essentially insulating and broad transport gaps (gray shaded regions) can be observed, if the leads
are attached only to these atoms. The confining zigzag edges of the studied finite armchair
nanoribbons stamp their fingerprint on the transport and can suppress the metallicity in armchair
ribbons.

The behavior of the LDOS in \fig{3} can be understood by the sublattice structure of
graphene.\cite{Brey2006, CastroNeto2009, Heikkilae2013} At zigzag edges, see for example in \fig{1}
(a) the edge to the left hand side, all inner atoms belong to one sublattice A (white resonators),
whereas all outer atoms belong to the other sublattice B (black resonators). If the nanoribbon is
extended hypothetically by one layer of atoms into the region, where the wave function has to vanish
due to hard-wall boundary conditions, all these atoms would belong to sublattice A only. Thus, at a
zigzag edge the wave function has to vanish on the sublattice to which also the inner atoms belong,
whereas no constraint has to be fulfilled by the wave function for the sublattice of the outer
atoms. This makes it possible that on the outer atoms of zigzag edges a edge (or surface) state can
reside, while the wave function vanishes essential on the inner atoms. Moreover, the surface states
are localized on one of the two $K$ points, which makes them robust against
perturbations.\cite{Fujita1996, Nakada1996, Brey2006, Wakabayashi2009, Wakabayashi2010,
  CastroNeto2009, Heikkilae2013} The localization of the edge states on the outer atoms of graphene
zigzag edges can be understood also from a simple resonance theoretic picture as well as from
valence bond and molecular orbit theory.\cite{Klein1999} These edge states have been observed in
microwave experiments.\cite{Bellec2014} Take into account that connecting leads to the inner atoms
on the zigzag edges is not equivalent to removing the outer atoms. This would lead to beard edges,
which have edge states and show different transport properties.\cite{Bellec2014} Note that we have
two different origins for the observed transport gaps depending on the width of the ribbons. While
for ribbons of width $3m$ or $3m+1$ the transport gap is directly related to a band gap in the
energy spectrum of the system, the engineering of atomically precise contacts allows to utilize the
spatial structure of the edge states and opens a transport gap in absence of a gap in the spectrum.

\subsection{Zigzag ribbons}
\label{sec:3-2}

The transmission through zigzag ribbons is shown in \fig{4} (top row). The leads are attached to the
inner and outer atoms of the confining armchair edges, see the black shaded sites of the nanoribbons
sketched in the insets. Experimental data are shown by blue-solid curves, calculations by red-dashed
curves.

As in \fig{2}, the transmission decreases when the Dirac point around $\nu=0$ is approached but it
remains finite for all ribbons. In agreement with theoretical predictions,\cite{Nakada1996,
  Brey2006, Wakabayashi2009, Wakabayashi2010, CastroNeto2009, Heikkilae2013} all zigzag ribbons
behave metallic independent from their actual width.

\begin{figure}
  \centering
  $\nu= -1 \un{MHz}$ \hspace{25mm}  $\nu= 5 \un{MHz}$\\
  \includegraphics[scale=0.42]{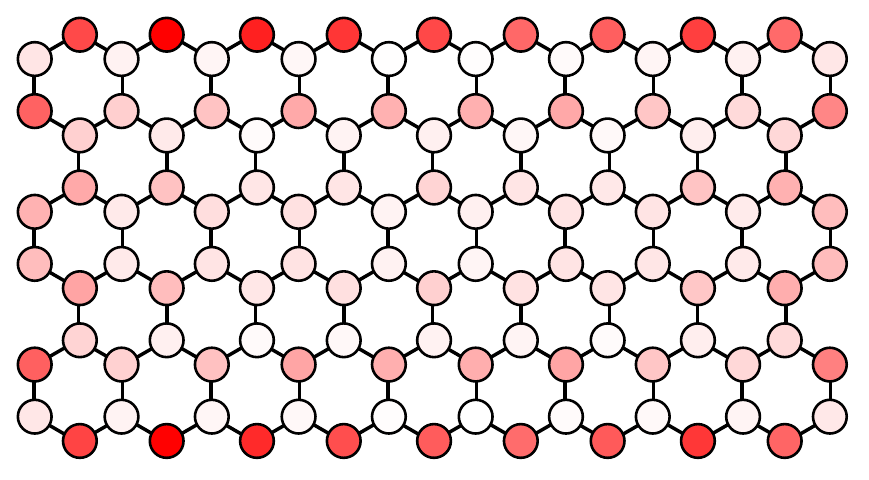}  \includegraphics[scale=0.5]{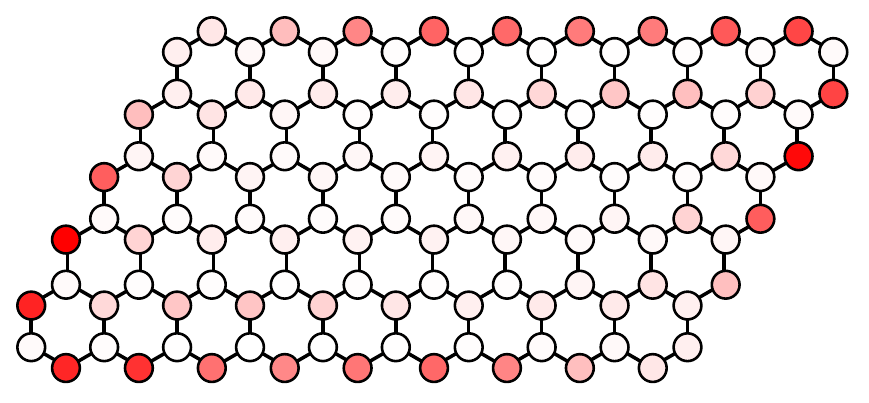}
  \caption{Left: In the zigzag ribbon the calculated LDOS is non-vanishing on all atoms of the
    confining armchair edges, where the leads are attached. A transport gap cannot be induced by
    contact engineering. Right: At the edges of the zigzag-rhomboid ribbon, the calculated LDOS is
    located on the outer atoms and vanishes on the inner atoms.}
  \label{fig:5}
\end{figure}

In zigzag ribbons a transport gap cannot be opened by contact engineering. The calculated LDOS in
\fig{5} (left) close to $\nu=0$ is finite for all atoms on the confining armchair edges to the left
and right hand side, because atoms form both sublattices appear there, see the alternating black and
white resonators in \fig{1} (b). A surface (or edge) state does not exist at armchair edges. This
behavior can be observed for all frequencies close to $\nu=0$. Moreover, solving the corresponding
Dirac equation, the wave function is located on both $K$ points (valley mixing), which makes it
sensitive to perturbations.\cite{Brey2006, CastroNeto2009, Heikkilae2013} Note that for the ribbons
of width $2$ and $3$, the experiment shows a small transport gap, which is up to now not understood
and not observed in our theoretical calculations.

In the zigzag-rhomboid ribbons, depicted for example in \fig{1} (c), all edges have the zigzag
shape. This allows to tune the transport in the system by suitable contact geometries. Close to the
Dirac point the calculated LDOS in \fig{5} (right) is located on the outer atoms of the zigzag edges
whereas it vanishes on the inner atoms. Therefore, if the leads are attached to the inner atoms of
the edges to the left and right hand side, see \fig{4} (bottom row), a broad transport gap is
observed. These gaps are observed for all ribbons independent from their size.

\section{Conclusions \& outlook}
\label{sec:4}

In this paper, the electronic transport in graphene nanoribbons has been studied by microwave
emulation experiments and tight-binding Green's function calculations. The microwave experiment
emulates only the ballistic single-particle transport. Correlations due to interactions between the
electrons, which may be present in real graphene, cannot be taken into account.

We have presented experimental evidence that the width of armchair ribbons determines whether the
ribbon is metallic or semiconducting, see \fig{2} (top row). We have also shown that all
(rectangular-shaped) zigzag ribbons are metallic, independent from their actual size, see \fig{4}
(top row).

We have found that the transport properties can be tuned by the contact geometry, using the fact
that the zigzag edge state resides on the outer atoms but not on the inner atoms, see the local
density of states in \fig{3} and \fig{5}. Hence, broad transport gaps can be induced in those
ribbons, where the leads are attached to the inner atoms of zigzag edges, see the armchair ribbons
in \fig{2} (bottom row) as well as the zigzag-rhomboid ribbons in \fig{4} (bottom row). The
realization of such contact geometries may be extremely difficult in real graphene. However,
considering the recent progress,\cite{Cai2010, Koch2012, Ruffieux2012, Chen2013b, Cai2014, Chen2015,
  Kimouche2015, Ruffieux2016} we think that this can be possible in the near future. Our microwave
emulation experiments thus may be viewed as a testing ground for new concepts.

In the future, we plan to investigate in more detail the transport in rhomboid-shaped ribbons, where
the contacts are attached to the corners. For example, it has been predicted
theoretically\cite{Shimomura2011, Cuong2013} that in zigzag-rhomboid ribbons at the Dirac point the
$60^\circ$ corners are conducting, while $120^\circ$ corners are insulating. This theory can be
confirmed by the local density of states in \fig{5} (right). However, first experiments show
surprisingly strong discrepancies to our tight-binding calculations, which we cannot explain at this
point and require further studies.

\begin{acknowledgments}
  We thank N. Szpak for fruitful discussions and his comments about the manuscript. Financial
  support from CONACyT research grant 219993 and PAPIIT-DGAPA-UNAM research grants IG100616 and
  IN114014 is acknowledged. T.S. acknowledges a postdoctoral fellowship from DGAPA-UNAM. T.H.S. and
  J.A.F.-V. are grateful for the hospitality regularly received at the LPMC.
\end{acknowledgments}

\bibliography{./grnrb}

\end{document}